\documentclass[sigconf]{acmart}

\usepackage{lipsum}
\usepackage{xcolor}
\usepackage{pifont}
\usepackage{diagbox}
\usepackage{url}
\usepackage{multicol}
\usepackage{multirow}
\usepackage{tabularx}
\usepackage{booktabs}
\usepackage{longtable}
\usepackage{footmisc}
\usepackage{amsmath}
\usepackage{cleveref}
\usepackage{balance}

\definecolor{green}{HTML}{009900}
\definecolor{orange}{HTML}{FF8000}
\definecolor{blue}{HTML}{6C8EBF}
\definecolor{dirtyyellow}{HTML}{D6B656}
\definecolor{g1}{HTML}{84a02b}
\definecolor{o1}{HTML}{ff9b21}
\definecolor{b1}{HTML}{7477bc}
\definecolor{y1}{HTML}{F8AE29}
\definecolor{so1}{HTML}{FF9D3A}
\definecolor{sg1}{HTML}{548135}

\AtBeginDocument{%
  }

\copyrightyear{2024}
\acmYear{2024}
\setcopyright{acmlicensed}
\acmConference[MM '24] {Proceedings of the 32nd ACM International Conference on Multimedia}{October 28--November 1, 2024}{Melbourne, VIC, Australia.}
\acmBooktitle{Proceedings of the 32nd ACM International Conference on Multimedia (MM '24), October 28--November 1, 2024, Melbourne, VIC, Australia}
\acmISBN{979-8-4007-0686-8/24/10}
\acmDOI{10.1145/3664647.3681472}

\begin{document}

\title[Auto-ACD]{Auto-ACD: A Large-scale Dataset for Audio-Language  \\ Representation Learning}

\author{Luoyi Sun}
\authornote{Work done during internship at Shanghai Jiao Tong University.}
\affiliation{%
  \institution{CMIC, Shanghai Jiao Tong University}
  \state{Shanghai}
  \country{China}
  }
\affiliation{%
  \institution{Shanghai Artificial Intelligence Laboratory}
  \state{Shanghai}
  \country{China}}
\email{loiesun411@gmail.com}

\author{Xuenan Xu}
\orcid{0000-0001-8718-1278}
\affiliation{%
  \institution{X-LANCE, Shanghai Jiao Tong University}
  \state{Shanghai}
  \country{China}}
\email{wsntxxn@sjtu.edu.cn}

\author{Mengyue Wu}
\authornote{Corresponding authors.}
\orcid{0000-0002-5599-8707}
\affiliation{%
  \institution{X-LANCE, Shanghai Jiao Tong University}
  \state{Shanghai}
  \country{China}}
\email{mengyuewu@sjtu.edu.cn}

\author{Weidi Xie}
\authornotemark[2]
\affiliation{%
  \institution{CMIC, Shanghai Jiao Tong University}
  \state{Shanghai}
  \country{China}}
\affiliation{%
  \institution{Shanghai Artificial Intelligence Laboratory}
  \state{Shanghai}
  \country{China}}
\email{weidi@sjtu.edu.cn}

\begin{abstract}
Recently, the AI community has made significant strides in developing powerful foundation models, driven by large-scale multimodal datasets. 
However, for audio representation learning, 
existing datasets suffer from limitations in the following aspects: 
insufficient volume, simplistic content, and arduous collection procedures. 
To establish an audio dataset with high-quality captions, we propose an innovative, automatic approach leveraging multimodal inputs, such as video frames, audio streams.
Specifically, we construct a \textit{large-scale, high-quality, audio-language dataset}, 
named as \textbf{Auto-ACD}, comprising over 1.5M audio-text pairs. 
We exploit a series of pre-trained models or APIs, 
to determine audio-visual synchronisation, 
generate image captions, object detection, or audio tags for specific videos. 
Subsequently, we employ LLM to paraphrase a congruent caption for each audio, guided by the extracted multi-modality clues. To demonstrate the effectiveness of the proposed dataset, 
we train widely used models on our dataset and show performance improvement on various downstream tasks, for example, 
audio-language retrieval, audio captioning, zero-shot classification.
In addition, we establish a novel benchmark with environmental information and provide a benchmark for audio-text tasks.   
\end{abstract}

\ccsdesc[500]{Computing methodologies~Computer vision}
\ccsdesc[500]{Information systems~Data structures}
\ccsdesc[500]{Information systems~Information retrieval}

\keywords{Audio-language Dataset, Audio-language Representation Learning, Audio Captioning}

\begin{teaserfigure}
  \includegraphics[width=\textwidth]{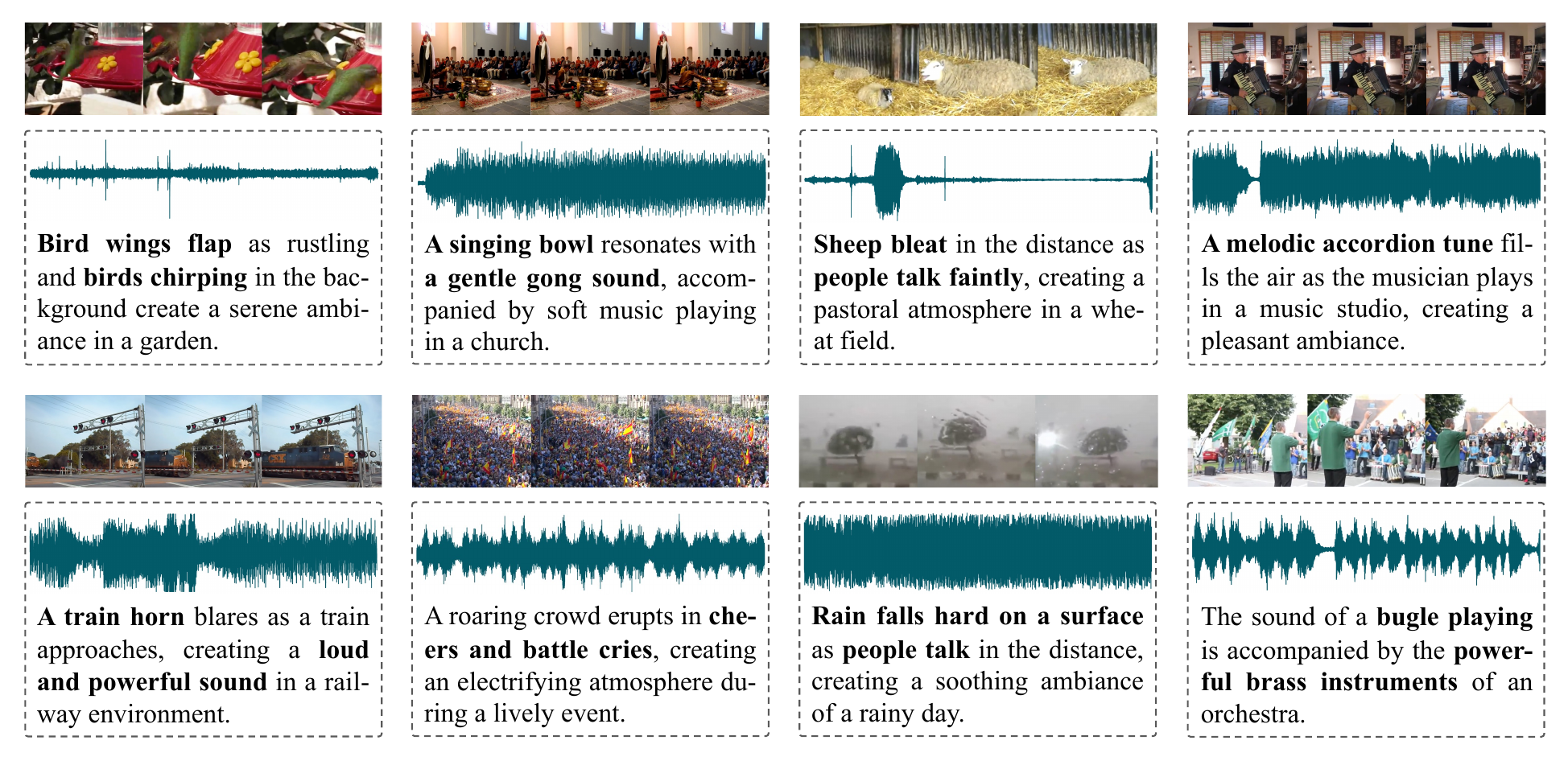}
  \caption{Examples from our proposed Auto-ACD, that is a large-scale audio-language dataset of audio-text pairs (1.5M), with long sentences (18 words) and diverse vocabularies (23K), 
  The captions for the audios in Auto-ACD are generated by an automatic pipeline, comprising elaborate sound descriptions, abundant auditory incidents and unique environmental information. The pivotal sound events are highlighted in bold. }
  \label{fig:teaser}
\end{teaserfigure}

\maketitle

\section{Introduction}

In recent literature, foundation models, like CLIP~\cite{radford2021learning}, variants of GPT~\cite{radford2019gpt2}, DALL-E 2~\cite{ramesh2022hierarchical} and Stable Diffusion~\cite{rombach2022high}, have shown tremendous success in various understanding and generation tasks. 
Despite being different in architectural or algorithmic designs, they are lying on a common basis: large-scale multimodal datasets, for example, MMC4~\cite{zhu2024multimodal}, LAION~\cite{schuhmann2022laion}, HowTo100M~\cite{miech2019howto100m}, indicating an emerging transition from model-centric to data-centric representation learning. The former considers pushing the boundaries of model design within the constraints of a predetermined data budget, while the latter focuses on curating large-scale and high-quality datasets in a scalable manner.

\begin{figure}[t]
    \centering
    \includegraphics[width=1\linewidth]{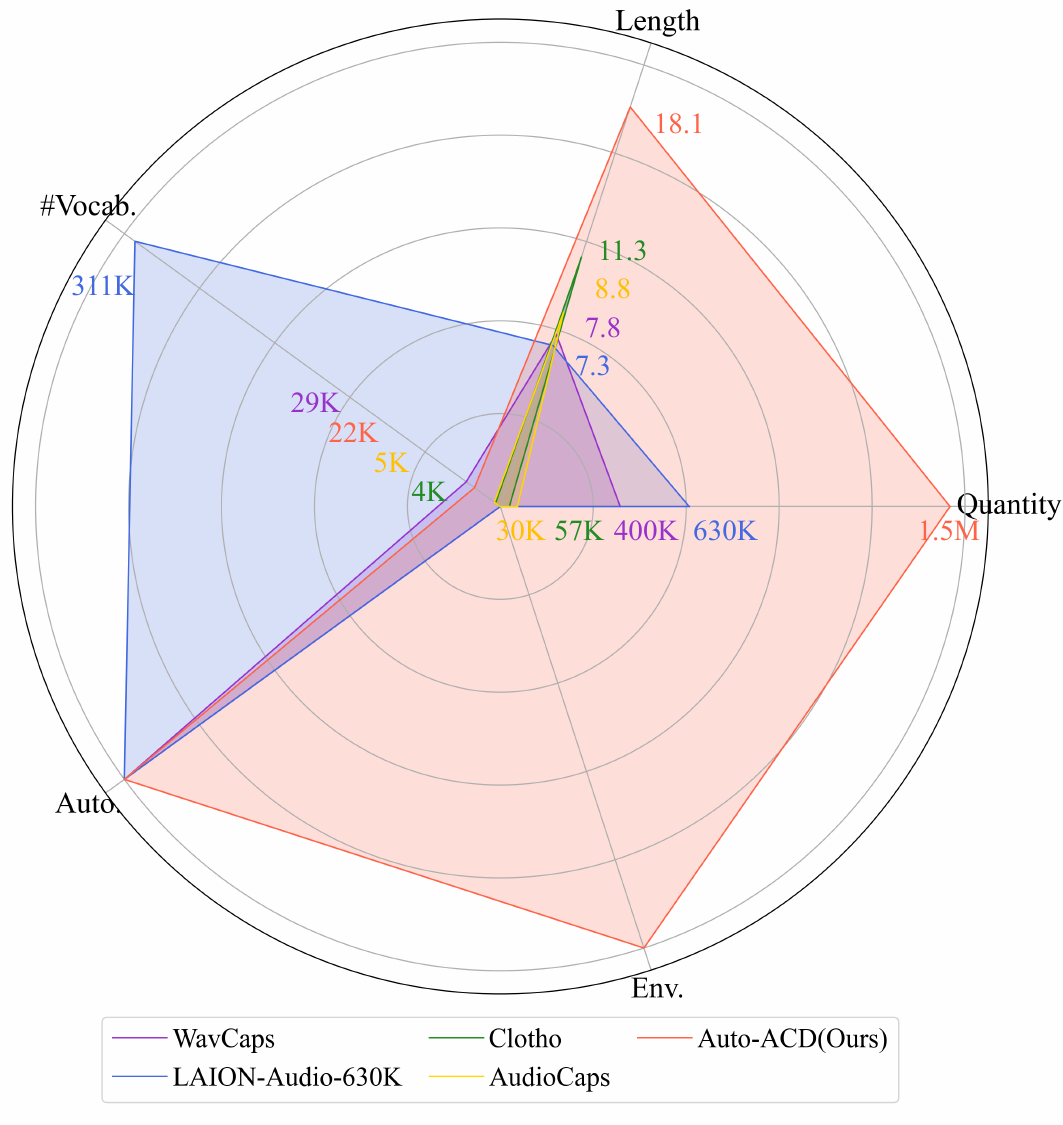}
    \caption{Comparison with other audio caption datasets. ``Length'' and ``\# Vocab.'' refer to average length and vocabulary. ``Env.'' and ``Auto.'' refer to environmental information and automatic pipeline, respectively.}
    \label{fig:radar}
\end{figure}

In the audio community, there have been recent endeavours on constructing audio-language datasets, as demonstrated in Fig.~\ref{fig:radar}. 
However, existing datasets potentially suffer from two limitations, laborious and complicated collection processes and simplistic descriptions in text. On the one hand, Clotho~\cite{drossos2020clotho} and AudioCaps~\cite{kim2019audiocaps}, which contain audios typically comprising 1 to 3 sound events, accompanied by high-quality text descriptions provided by human annotators. This is are clearly challenging to scale up. On the other hand, LAION-Audio-630K~\cite{wu2023large} and WavCaps~\cite{mei2023wavcaps} collect large amounts of raw data from online foley websites, then employ sentence templates or keyword-to-caption models to convert the original audio labels into free-form sentences. It is obvious that the resulting language descriptions hardly offer additional information than simple prompts or sound tags. 
Therefore, models trained on these datasets are incapable of learning robust audio-language representations.

This paper presents our recent efforts for constructing a large-scale, high-quality, audio-language dataset, with minimal manual efforts, termed \textbf{Auto-ACD} (\textbf{A}udio \textbf{C}aptioning \textbf{D}ataset by \textbf{Auto}matic Collection), with massive audio-text pairs (1.5M), long texts (18 words) and diverse vocabularies (23K).
Specifically, an exemplary audio caption ought to encapsulate four varieties of information: the `what' - the nature of the sound perceived, the `who' - the entity producing the sound, the `how' - the characteristics of the sound, and the `where' - the location the sound occurs. 

Our key insight is that comprehensive understanding of the visual scene is expected to serve as a valuable information source and is sometimes necessary for understanding the audio content. Therefore, we build Auto-ACD on the prior of robust audio-visual correspondence in existing audio-visual datasets, for example, VGGSound~\cite{chen2020vggsound}, AudioSet~\cite{gemmeke2017audio}. Particularly, we initiate an automatic pipeline, that employs a range of publicly available tools or APIs across the general AI community, {\em e.g.}, vision, language and audio models, to generate comprehensive language descriptions for the audio stream of the given video datasets. Lastly, we employ a large language model (LLM) to collectively assimilate all outputs, identify and eliminate any illogical information, and generate comprehensive descriptions for the audio. As a result, these descriptions not only depict the type of sound and its source, but also describe the auditory attributes and the specific location of its occurrence.


To comprehensively validate auditory representation, for instance, audio events, and ambient information, learned from the text descriptions of Auto-ACD, we conduct experiments from four perspectives:
{\em First}, we launch a joint audio-language representation learning using InfoNCE loss~\cite{oord2018representation, han2019video, zhou2023adaptive}, and evaluate the model through a retrieval task between audio and language, showing noticeable improvement over existing datasets; {\em Second}, we conduct zero-shot classification experiments, demonstrating the effectiveness for learning environmental information with our dataset; {\em Third}, we benchmark on audio-language generation task, specifically, automatic audio captioning, by training a lightweight mapping network between the pre-trained audio backbone and GPT2~\cite{radford2019gpt2}, showing superior performance on the widely used benchmark, {\em e.g.}, Clotho~\cite{drossos2020clotho};
{\em Fourth}, we manually filter a test set and introduce a novel benchmark for audio-language tasks. This benchmark assesses the ability of models to grasp information beyond mere audio tags, for example, the environment and fine-grained categories of sound, we set a baseline for future research in this field.

\begin{figure*}[t]
  \centering
  \includegraphics[width=\textwidth]{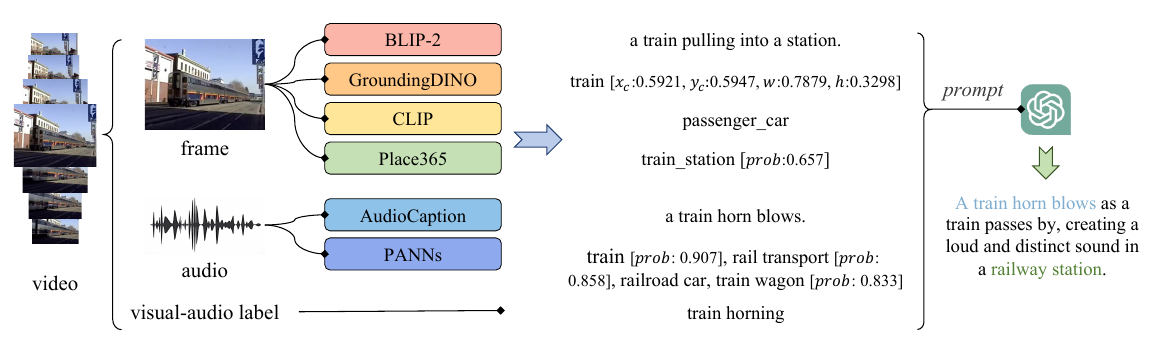}
  \caption{Automatic pipeline for Auto-ACD collection. We utilize four open-source computer vision models to extract visual clues from the middle frame of videos, and two open-source audio understanding models to analyze the entirety of the audio content. Consequently, we combine the labels from the original dataset, and leverage Large Language Models (LLMs) to interpret and paraphrase these components into the final description.}
  \label{fig:pipeline}
\end{figure*}
\section{Related Work}

\subsection{Audio-visual Learning}
Audio-visual events often occur simultaneously within in-the-wild videos, establishing a profound connection between sound and imagery.
\cite{arandjelovic2017look, alayrac2020self, jenni2023audio, owens2018audio} employ audio-visual self-supervised learning to leverage audio-visual correspondence for enhancing representation learning. Specifically, \cite{guzhov2022audioclip,wu2022wav2clip,zhao2022connecting} learn audio-text representation based on such correspondence.
Audio-visual localisation \cite{chen2021localizing, sun2023learning, mo2022closer, hu2022mix, mo2023audio} concentrates on identifying the positions of visual sound sources within video.
Audio-visual segmentation \cite{zhou2022audio, liu2024annotation, gao2024avsegformer, li2023catr, mo2023av} 
aims to predict the pixel-wise segmentation masks of sounding objects in visual scenes precisely. Such studies have further demonstrated the intrinsic correlation between audio and visual events in in-the-wild videos, which inspires us to create an audio-language dataset anchored in visual information.

\subsection{Audio-visual Dataset}
Large-scale audio-visual datasets are crucial for effective audio and video understanding. Two datasets are often involved in audio-visual learning: AudioSet and VGGSound.
AudioSet~\cite{gemmeke2017audio} is a large-scale audio-visual dataset with multiple audio events labelled for each audio clip.
It contains over 2M 10-second audio clips.
AudioSet is a manually annotated dataset, with the help of a well-structured hierarchical ontology consisting of 632 audio classes guided by literature and manual curation.
VGGSound~\cite{chen2020vggsound} comprises 200K 10-second videos for 309 audio classes. This dataset was collected and annotated through an automated pipeline, with each video assigned only one label.
In this paper, we aim to provide detailed description for audios, by exploiting both audio and visual cues.

\subsection{Audio-language Learning}
The application of visual-language models in the audio-language arena marks a significant leap forward. Notably, \cite{wu2023large} have adapted the CLIP model for audio-language contrastive learning, setting a precedent for innovative cross-modal research. 
Researchers are not merely focusing on extracting semantic information from audio through tasks such as audio-text retrieval~\cite{oncescu2021audio, koepke2022audio}, audio classification~\cite{hershey2017cnn, palanisamy2020rethinking}, automatic audio captioning~\cite{xu2022comprehensive, mei2022automated}, and audio question answering~\cite{lipping2022clotho,li2023multi}.
They are also venturing into more nuanced aspects of auditory perception, including exploring temporal dynamics in sound through audio event detection \cite{babaee2017overview,li2023ast}.
This broadening scope encompasses additional auditory attributes such as counting sounds within scenes~\cite{nigro2023sardbscene} and classifying environments based on their acoustic characteristics\cite{ding2023acoustic}.
Undoubtedly, it is paramount to construct a comprehensive, large-scale, high-quality and information-rich audio-language dataset.

\subsection{Audio-language Dataset}
Audio-language tasks, including audio-text retrieval, audio captioning, audio question answering and text-guided audio generation, have greatly benefited from the availability of two widely-used audio captioning datasets: AudioCaps and Clotho.
AudioCaps~\cite{kim2019audiocaps}, a subset of AudioSet, consists of 50K 10-second-long audio clips, each with a single caption annotated.
The annotators were provided with AudioSet tags as hints and videos if necessary.
Clotho~\cite{drossos2020clotho}, on the other hand, comprises 6K audio clips lasting between 15 to 30 seconds, each with five captions annotated through a three-step process involving captioning, grammar correction, and rating by human annotators. 
However, due to the human annotation process, these datasets are limited in size, expensive and time-consuming.
LAION-Audio-630K~\cite{wu2023large} acquires audio and descriptions from online foley websites, including popular platforms like Freesound\footnote{https://freesound.org/} and BBC Sound Effects\footnote{https://sound-effects.bbcrewind.co.uk/}. 
WavCaps~\cite{mei2023wavcaps} utilizes ChatGPT to filter and paraphrase these raw descriptions, resulting in a dataset of 400K audio-text pairs with cleaned text data resembling human annotations. 
The sentence is mostly simple since there is often only one sound event in an audio clip. 
As a result, models trained on these datasets could only learn the category of sound. To enhance the comprehension capabilities of the audio-text model, we need a more diverse set of textual and audio data.

\section{Dataset Construction}

To develop a large-scale, audio dataset with rich language descriptions, we base on the assumption that visual scene understanding serves as a strong prior. 
For instance, synchronized videos frequently showcase auditory cues, and visual information serves as a precise representation of the acoustic environment in which the sound happens. 

In an audio caption, it is desirable to incorporate sound attributes, location, and fine-grained labels. 
To achieve this, we can leverage publicly available tools or APIs to gather the necessary information for audio description and mutually verify the results. 
For instance, we can employ an object detection model to identify potential sources of sound, and an environmental classification model to extract scene categories. 
By extracting a wealth of information, we ensure the maximum coverage of accurate details, providing the language model with ample references. 

\subsection{Tools or APIs}

Given one sample from existing large-scale video datasets, 
for example, AudioSet or VGGSound~\cite{chen2020vggsound,gemmeke2017audio}, 
{\em i.e.}, denoted as $\mathcal{V} = \{f; a; y\}$, 
where $f$, $a$ and $y$ correspond to frame sequence, 
audio stream, and visual or audio labels, respectively.
Our goal is to adopt a range of publicly available tools or APIs across the general AI community, {\em i.e.}, using off-the-shelf vision, language and audio models to construct language descriptions for audios, as shown in Fig.~\ref{fig:pipeline}.
In this section, we describe these tools in detail.

\subsubsection{Image Captioning.}
We employ the off-the-shelf BLIP-2~\cite{li2023blip} model, 
which obtains competitive results for image captioning.
This tool has the ability to generate captions 
that encompass the entire image and accurately depict the primary subject or environment. In our case, we input the middle frame of the video into this model.

\subsubsection{Object Detection.}
We use the pre-trained Grounding DINO model~\cite{liu2023grounding}, 
to identify objects within the middle frame, and preserve all the detected entities along with their corresponding prediction confidence scores to ensure a comprehensive analysis. 
 
\subsubsection{Image Labeling.}
We adopt the pre-trained OpenAI CLIP~\cite{radford2021learning} model for image classification.
In this application, we utilize the prompt: ``a photo of a \{label\}" to generate textual embedding, 
leveraging the category ontology from ImageNet~\cite{deng2009imagenet}.

\subsubsection{Place Recognition.}
We employ the pre-trained PlaceCNN~\cite{zhou2017places}, to infer the environment context captured in videos. 
Given the robust correspondence between audio and visual signals, 
the environment depicted in the video is highly likely to represent the acoustic ambience in which the sound occurs.

\subsubsection{Audio Tagging.}
We use the pre-trained PANNs~\cite{kong2020panns} to predict the tags of sounds within the audio, and preserve the top three predictions with their confidence scores. This represents a crucial source of auditory temporal information, particularly for sounds emanating from entities not visible within the frame.

\subsubsection{Audio Captioning.}
We use the existing AudioCaption~\cite{xu2023blat} model, to generate concise and brief captions. These captions resemble the style of AudioCaps, focusing solely on the categorical information of audio events, devoid of any additional descriptive attributes about the sound.

\subsubsection{Audio-visual Synchonisation.}
We employ the pre-trained Synchformer~\cite{iashin2024synchformer} to conduct synchronization detection between video and audio. This process could filter out samples consisting of irrelevant or unsynchronized video and audio content. In this case, we input both video and audio respectively into this model for analysis.

\subsubsection{Existing Audio-Visual Labels.}
In addition to the predictions from models, 
we also incorporate the provided labels of existing datasets into our pipeline. For instance, VGGSound~\cite{chen2020vggsound} gives a single label for each video,
while AudioSet~\cite{gemmeke2017audio} provides multiple labels.
These labels serve in the original dataset, offering accurate yet incomplete audio-visual information.

\begin{figure}[t]
  \centering
  \includegraphics[width=0.48\textwidth]{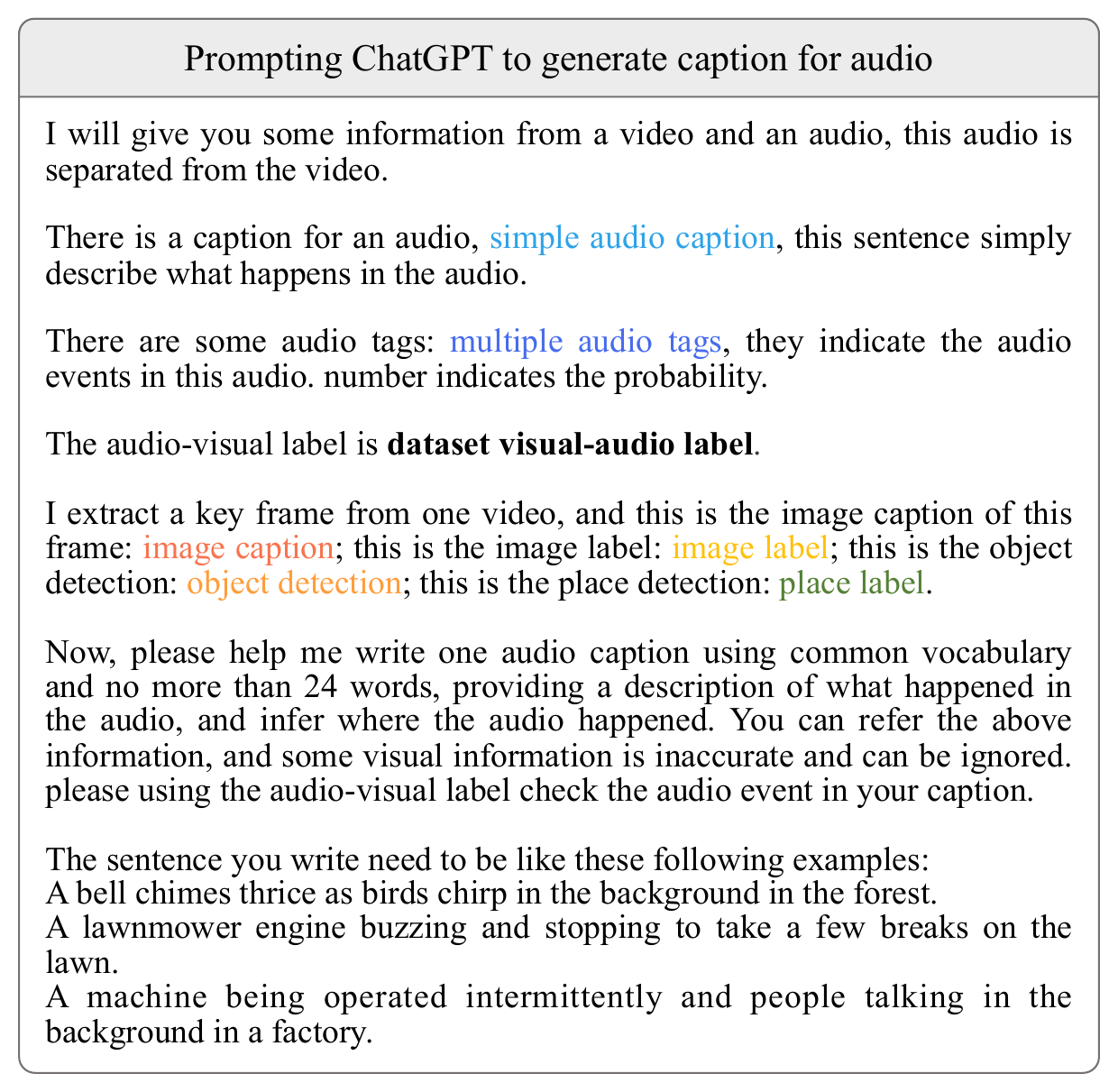}
  \caption{Detailed prompt provided to ChatGPT. For visualisation purposes, we use different colors to highlight diverse visual-audio cues.}
  \label{fig:prompt}
\end{figure}

\subsubsection{Summary.}
As for the language model, we use the OpenAI ChatGPT\footnote{https://openai.com/chatgpt}, which demonstrates strong performance in reasoning and inductive summarization, 
to assemble the above-mentioned descriptions or labels into comprehensive descriptions for audio.
Many works, like BLIP-2\cite{li2023blip}, show that utilizing existing tools appropriately can significantly enhance the model's performance. By leveraging audio-visual correspondence and the profound understanding capabilities of LLM, we generate precise audio captioning from the rich multi-modality clues acquired.
In this case, we feed in a special prompt as shown in \Cref{subsec:caption_generation}.

\subsection{Caption Generation}
\label{subsec:caption_generation}
Based on the visual and acoustic clues present in the video, we craft a structured language paragraph, and use it to prompt ChatGPT to generate descriptions for audio. As illustrated in Fig.~\ref{fig:prompt}, the process begins with formulating the specific task and criteria for the desired outcome, followed by inputting seven distinctive audio-visual cues into the prompt, accompanied by their corresponding confidence score.
Additionally, we provide three sentence examples from AudioCaps or Clotho as instruction.
For visualisation purposes, we here use a colour-coded system to distinguish various cues.

While generating captions, we explicitly ask ChatGPT to remove information that is inaudible, {\em i.e.}, illogical and visually oriented elements, for example, colours. As a result, the large language model is able to analyze the scenario from all provided clues, and generate language description for audio, with sound category, and environment. The generated caption results are shown in Table.~\ref{tab:visualization}.

\subsection{Dataset Filtering}
\label{sec:filter}
\begin{figure}[t]
    \centering
    \includegraphics[width=\linewidth]{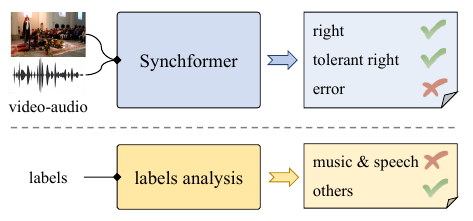}
    \caption{Filtering process for AudioSet. We filter the dataset by assessing whether the video and audio are synchronized and analyzing the labels in the original dataset.}
    \label{fig:filter}
\end{figure}

AudioSet is vast and diverse, while heavily marred by noise in many instances, for instance, gameplay live streams and explanatory videos.
Conversely, VGGSound significantly emphasises the robust correlation between video and audio within the automated collection pipeline, thus requiring no further processing. As shown in Figure.~\ref{fig:filter}, we formulate filtering criteria grounded in both the video-audio correspondence and the original labels.
For each filter criterion, we conduct numerous trials followed by a manual verification, each filtering criterion achieves an accuracy rate exceeding 90\%, resulting in the removal of 0.4 million videos in total. 

\begin{table}[t]
    \caption{The results of generated captions in Auto-ACD, with accurate content and ample surrounding information. \textcolor{g1}{Green} and \textcolor{y1}{Yellow} refer to ``where" and ``how" the audio sounds like.}
    \label{tab:visualization}
    \setlength{\tabcolsep}{2pt} 
    \centering
    \begin{tabular}{ll}
    \toprule
     No. & Generated Caption \\
    \midrule
    
    \multirow{2}{*}{1.} & Loud pops and bangs resonate as timbales are being played,  \\
     & creating \textcolor{y1}{rhythmic} music \textcolor{g1}{in a room}. \\
    \midrule
    
    \multirow{2}{*}{2.} & Water gurgles and bubbles as a boat glides through, creating \\
      & a \textcolor{y1}{soothing and peaceful} \textcolor{g1}{underwater} ambience.\\
    \midrule
    
    \multirow{2}{*}{3.} & A woman speaks softly amidst the soothing sound of birds  \\
      &  chirping, creating a \textcolor{y1}{serene} atmosphere \textcolor{g1}{in a garden}.\\
    \midrule
    
    \multirow{2}{*}{4.} & A motorcycle engine idles before revving up,  creating a \textcolor{y1}{loud} \\
      &  sound \textcolor{g1}{in an urban environment}.\\
      
    \bottomrule
    \end{tabular}
\end{table}

\subsubsection{Raw labels.}
AudioSet contains a plethora of explanatory videos with background music, wherein the visual and auditory information often do not correspond. Therefore, we eliminate videos from the multi-labels that encompass both speech and music.

\subsubsection{Audio-visual synchronisation.}
To obviate the possibility of fortuitous inference errors, we subject each video to five synchronization evaluations, featuring random variations in start time and offset, with a tolerance threshold established at 0.6 seconds. 
Synchformer\cite{iashin2024synchformer} employs a 0.2s offset to ascertain the precise audio-visual synchronization, whereas we utilize a broader offset to determine the audio-visual correspondence.
The outcomes are categorized as follows: (1) Predictions aligning with the ground truth are deemed ``correct''; (2) Predictions that diverge from the ground truth while with a discrepancy within 0.6 seconds are designated as to be ``tolerable''; (3) All other results are termed ``error''. To preserve as much data as possible, videos classified as ``error'' in all five tests are removed from the dataset.

\begin{figure*}[t]
    \centering
    \includegraphics[width=\linewidth]{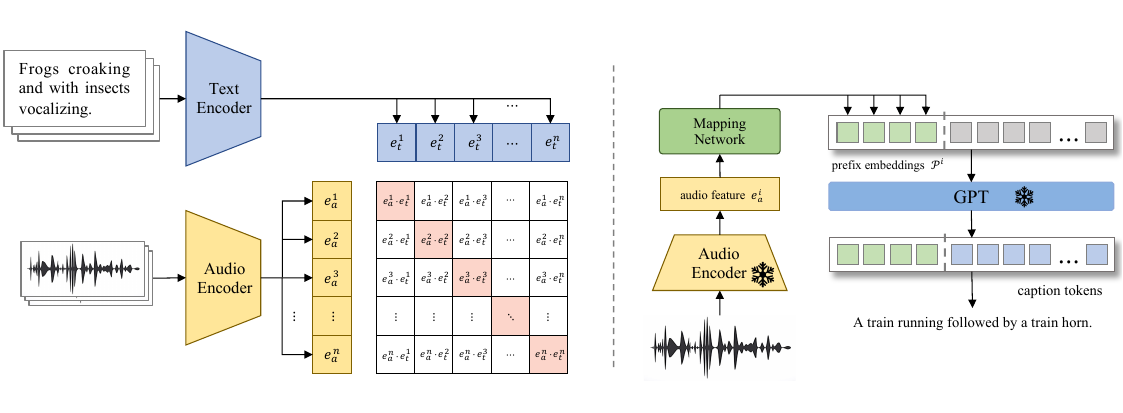}
    \caption{Audio-language retrieval model and automatic audio captioning model frameworks. Similar to CLIP, the audio-language retrieval model consists of an audio encoder, text encoder, and contrastive loss. The automatic audio captioning model comprises a frozen audio encoder and language model, and a trainable mapping network. }
    \label{fig:architecture}
\end{figure*}

\subsection{Dataset Statistics}
As depicted in Fig.~\ref{fig:radar}, 
we collect 1.5 million audio-language pairs from AudioSet and VGGSound in total.
To the best of our knowledge, Auto-ACD is the first million-level audio-language dataset to date,
with train, validation and manually filtered test sets.
Auto-ACD surpasses the other datasets in terms of data volume, average sentence length, and contains a relatively wide verbal vocabulary. 
LAION-Audio-630K\cite{wu2023large} sources from user uploads, contains a plethora of noisy details, for instance, device and timestamps, and features an exceptionally extensive vocabulary.  
Additionally, Auto-ACD stands as the only audio-language dataset that encompasses environmental information, not only delineates the type and source of sounds but also specifies the location of their occurrence, increasing the richness of contextual details.

In supplementary, we present a comparative analysis of captions from LAION-Audio-630K, WavCaps, and Auto-ACD for the same audio sample.
Captions in LAION-Audio-630K and WavCaps are concise and contain minimal information beyond the audio tags.
In particular, LAION-Audio-630K may include sentences that deviate from common sense, for example, describing ``rapping a tree'' for an audio tag of ``rapping''.
WavCaps, on the other hand, exhibits a monotonous sentence structure, such as ``... sound can be heard''.
In contrast, Auto-ACD features longer sentences that provide a richer depiction of the audio scenes.

We conduct a manual check on randomly sampled 200 audio-captions pairs from Auto-ACD, analyzing the clues from the different open-source tools and the generated captions. We define a clue that contradicts the audio to be erroneous, these tools possess high accuracy, the average accuracy is 81.3\%. Furthermore, we conduct a manual check on randomly sampled 1000 audio-captions pairs, and find that 92.4\% captions correspond with audio, just 5.3\% incorrect words need to be modified, and only 4.4\% captions contain inaudible information. These results indicate that our proposed approach enables high-quality, scalable caption generations, with few incorrect or inaudible information.

\section{Architecture}
We construct architectures targeting two general audio-language tasks, namely, audio-language contrastive learning, 
and automatic audio captioning, to further validate the effectiveness of Auto-ACD. In Section \ref{sec:retrieval}, we provide a detailed exposition of the architecture for audio-language contrastive learning. Further in Section \ref{sec:captioning}, we introduce the framework for lightweight automatic audio captioning along with its loss function.

\subsection{Audio-Language Constrastive Learning}
\label{sec:retrieval}
To validate the efficacy of our proposed dataset, 
we train an audio-language model with standard contrastive learning, 
{\em e.g.}, infoNCE~\cite{radford2021learning} loss, as shown in Fig.\ref{fig:architecture}.
Specifically, we employ the pre-trained HTSAT~\cite{chen2022hts} as the audio encoder, and the pre-trained RoBERTa~\cite{liu2019roberta} as the language encoder. Both encoders were initialised from the pre-trained CLAP model~\cite{wu2023large}, and further finetuned on our dataset.
We term our final model as Audio-Text Retrieval~(ATR).

Given an audio-text pair $(a^i,t^i)$, we utilise audio encoder $\mathcal{A}_{\text{enc}}$ and text encoder $\mathcal{T}_{\text{enc}}$ to extract audio embedding $e_{a}^{i}$ and text embedding $e_{t}^{i}$, respectively:
$$e_{a}^{i} = \mathcal{A}_{\text{enc}}(a^i), \text{\hspace{5pt}} e_{t}^{i} = \mathcal{T}_{\text{enc}}(t^i)$$
The model is then trained with contrastive loss, 
wherein the paired audio and language embeddings are treated as positive,
and unpaired ones as negative, with the following loss function:

\begin{equation*}
\mathcal{L} = \frac{1}{2 N} \sum_{i=1}^{N}  (\log \frac{\exp \frac{e_{a}^{i} \cdot e_{t}^{i}}{\tau} }{\sum_{j=1}^{N} \exp \frac{e_{a}^{i} \cdot e_{t}^{j}}{\tau}} +\log \frac{\exp \frac{e_{t}^{i} \cdot e_{a}^{i}}{\tau} }{\sum_{j=1}^{N} \exp \frac{e_{t}^{i} \cdot e_{a}^{j}}{\tau} })
\end{equation*}
where $\tau$ represents the learnable temperature parameters.

At training phase, we introduced word-level text masking, that is to randomly mask words within the sentences, before feeding into the text encoder.

\begin{table*}[htbp]
    \caption{The audio-text retrieval results on AudioCaps, Clotho and ACD test sets. ``basic'', ``LA.'' ``Wav.'' and ``ACD" refer to the combination of AudioCaps and Clotho~(basic), LAION-Audio-630K~(LA), WavCaps~(Wav) and Auto-ACD~(ACD), respectively. ``$\text{ACD}_\text{VS}$'' is a subset of Auto-ACD, curated from VGGSound. `` * FT'' refers to fine-tuning the model on the target dataset.}
    \label{tab:retrieval}
    \centering
    \setlength{\tabcolsep}{5pt} 
    \begin{tabular}{llcccccccccccc}
    \toprule
    \multirow{3}*{Train Set} & \multirow{3}*{Model} & \multicolumn{4}{c}{AudioCaps Test}  & \multicolumn{4}{c}{Clotho Test}  & \multicolumn{4}{c}{Auto-ACD Test} \\ 
  
    \cmidrule{3-14}
    &  & \multicolumn{2}{c}{Audio$\rightarrow$Text} & \multicolumn{2}{c}{Text$\rightarrow$Audio} & \multicolumn{2}{c}{Audio$\rightarrow$Text} & \multicolumn{2}{c}{Text$\rightarrow$Audio} & \multicolumn{2}{c}{Audio$\rightarrow$Text} & \multicolumn{2}{c}{Text$\rightarrow$Audio}\\
  
    &  &  \multicolumn{1}{c}{R@1} & \multicolumn{1}{c}{R@10} & \multicolumn{1}{c}{R@1} & \multicolumn{1}{c}{R@10} & \multicolumn{1}{c}{R@1} & \multicolumn{1}{c}{R@10} & \multicolumn{1}{c}{R@1} & \multicolumn{1}{c}{R@10} & \multicolumn{1}{c}{R@1} & \multicolumn{1}{c}{R@10} & \multicolumn{1}{c}{R@1} & \multicolumn{1}{c}{R@10}\\ 
  
    \midrule
    basic+LA.\cite{wu2023large} &  HTSAT-RoBERTa &45.0 & 88.0 & 36.2 & 82.5 & 24.2 & 66.9 & 17.2 & 55.4 & 20.0 & 65.0 & 17.9 & 59.7 \\
    basic+Wav.\cite{mei2023wavcaps} & HTSAT-BERT & 51.7 & 90.6 & 39.7 & 86.1 & 23.4 & 63.4 & 19.5 & 58.2  & - & - & - & -\\
    \midrule
    
    basic+$\text{ACD}_\text{VS}$ & HTSAT-RoBERTa & 50.5 & 90.6 & 39.8 & 86.9 & 24.2 & 62.9 & 20.0 & 58.9 & 39.2 & 86.2 & 39.6 & 85.7\\
    basic+ACD & HTSAT-RoBERTa &  53.7  & 91.7 & 39.5 & 85.4 & 17.7 & 52.6  & 15.3  & 52.1  & \textbf{47.1} & \textbf{91.2} & \textbf{49.0} & \textbf{92.3}\\
    basic+ACD*FT & HTSAT-RoBERTa & \textbf{56.3} & \textbf{93.9} & \textbf{42.7}  & \textbf{88.5} & \textbf{26.2} & \textbf{67.5} & \textbf{21.7} & \textbf{61.7}  & - & - & - & -\\
    \bottomrule
    \end{tabular}
\end{table*}

\subsection{Automatic Audio Captioning}
\label{sec:captioning}
To demonstrate the effectiveness of our pre-trained audio backbone, we also use audio captioning for evaluation.
Inspired by ClipCap~\cite{mokady2021clipcap} and AutoADs~\cite{han2023autoad1,han2023autoad2}, 
we adopt a lightweight audio captioning model, where both the audio backbone and language model~(GPT-2) are fixed, and only a mapping network is trained, 
as shown in Fig.~\ref{fig:architecture}.

Given an audio-text pair $(a^i,c^i)$, we use the pre-trained audio encoder to extract audio features $e_{a}^{i} =\mathcal{A}_{\text{enc}}(a^i)$, 
and we convert the caption into a token sequence, $c_{1}^{i}, \ldots, c_{k}^{i}$, where $k$ indicates the maximal length of text. Then, we design a mapping network $f_{\text{map}}$ to transform the extracted embedding into a set of prefix embeddings:
$$\mathcal{P}^i = f_{\text{map}}(e_{a}^{i}).$$

We take the prefix embedding set as the condition for predicting the next token with an auto-regressive language model. 
Therefore, during training, we minimize the negative log-likelihood of predicting the correct word:
$$
\mathcal{L}=-\sum_{i=1}^{N} \sum_{j=1}^{\ell} \log p_{\theta}\left(c_{j}^{i} \mid \mathcal{P}^i, c_{1}^{i}, \ldots, c_{j-1}^{i}\right)
$$
where $\theta$ represents the trainable parameters.

\section{Experiments}
In this section, we evaluate three tasks, namely, audio-language retrieval, audio captioning and zero-shot classification.

\subsection{Audio-language Retrieval}
\label{subsec:retrieval}

\subsubsection{Dataset.} 
We conduct audio-text retrieval experiments across several datasets: AudioCaps, Clotho, $\text{Auto-ACD}_\text{VS}$, and Auto-ACD. The distributions for the train, validation, and test sets in AudioCaps, Clotho, and Auto-ACD are 50K/495/975, 3.8K/1045/1045, and 1.5M/2K/1K data pairs, respectively.
$\text{Auto-ACD}_\text{VS}$ is a subset of Auto-ACD, 
containing 190K data pairs exclusively sourced from VGGSound. Notably, in the case of Clotho, and AudioCaps~(validation and test set), each data pair consists of one audio sample accompanied by five corresponding captions, while the remaining data pairs only comprise one audio-caption pair. 

\subsubsection{Auto-ACD Benchmark.}
In addition to the Auto-ACD training set, we also randomly selected 2K data samples to form the validation set and 1K samples for the test set. We conduct a \textbf{manual} verification of the test set, by removing incorrect information from the language descriptions and rewriting inappropriate vocabulary expressions. This test set is used for evaluating both audio-language retrieval and automatic audio captioning tasks.

\subsubsection{Metrics.}
In order to validate the rich and accurate information of our dataset, we compare the traditional metrics, Recall@$k$ performance, on commonly used datasets,
for example, AudioCaps and Clotho. 
We also adopt these metrics on the Auto-ACD test set, 
offering a comprehensive overview. 

\subsubsection{Training Details.}
We train our proposed Audio-Text Retrieval (ATR) model for 20 epochs, employing a batch size of 768, and utilizing the Adam optimizer with a warm-up phase, and an initial learning rate of 1e-4 with a cosine learning rate decay schedule. We use the same hyperparameters as those in the existing CLAP model configuration. The dimensions of both the audio encoder and text encoder output are 512.
Additionally, we introduce 25\% random masking on words in the sentences and randomly apply augmentations such as Noise and Gain to 50\% of audio samples to enhance the model training. We further fine-tune the model on specific datasets, for example, Clotho and AudioCaps, with an initial learning rate of 2e-5 for 15 epochs.

\subsubsection{Results.}
As shown in Table.\ref{tab:retrieval}, we can draw the following key observations: 
(i) comparing with training on Laion-Audio-630K, training on our proposed $\text{Auto-ACD}_\text{VS}$ dataset leads to a significant improvement in Recall@$k$ metrics on AudioCaps and Auto-ACD benchmarks. 
(ii) training on Auto-ACD and fine-tuning on specific datasets, which is not applicable to Auto-ACD benchmark, results in a remarkable performance gain.
This improvement is particularly evident when evaluating the model on the test set of AudioCaps, as AudioCaps is a subset of AudioSet and shares a similar data distribution with Auto-ACD. Such fine-tuning processes enable the model to acquire a more comprehensive understanding of both audio and text information, thus enhancing retrieval performance.
(iii) on the Auto-ACD benchmark, characterized by a more diverse lexicon and abundant language description, training on Auto-ACD datasets significantly outperforms the model trained on Laion-Audio-630K.

\subsection{Automatic Audio Captioning}

\subsubsection{Dataset.}
In addition to the datasets mentioned in \Cref{subsec:retrieval}, 
we also use the MACS dataset~\cite{martin2021diversity}, 
which comprises 3.9K audio-text data pairs,
with each audio accompanied by two to five captions and several audio tags.
In total, we train the automatic audio captioning model utilizing a total of 58k data pairs from Clotho, AudioCaps and MACS, and evaluate on Clotho and Auto-ACD test set.

\subsubsection{Metrics.}
In addition to conventional captioning metrics, for example, Meteor~\cite{banerjee2005meteor}, RougeL~\cite{lin2003automatic}, Spider~\cite{liu2017improved}, we incorporate SentenceBERT~\cite{reimers2019sentence} as additional evaluation metrics, that not solely rely on lexical alignment, but rather prioritize the semantic resemblance and accuracy of the captions' content.

\subsubsection{Training Details.}
We devise two mapping networks, MLP and transformer, 
and fine-tune the parameters of GPT during the training process.
We set the number of prefixes to be 8, each with a dimension of 512.
We train this audio captioning model on the MACS~\cite{martin2021diversity}, Clotho and AudioCaps for 15 epochs with a batch size of 128 and an initial learning rate of 5e-4. In this task, we compare the audio encoder from our pre-trained audio-text retrieval model and the pre-trained CLAP~\cite{wu2023large}, by only training the mapping network of both models on the benchmark datasets, namely, Clotho, and Auto-ACD.

\subsubsection{Results.}
As shown in Table.~\ref{tab:captioning}, we can draw two observations: 
(i) the automatic audio captioning model, with the audio encoder initialised from our pre-trained audio-text retrieval model, shows improved performance across all evaluation metrics than baseline.
(ii) there is a more pronounced outcome when evaluated on Auto-ACD: 
the baseline approach's performance oversees a sharp decrease in the test set of Auto-ACD.
We conjecture this is because the baseline features extracted from the CLAP model lack detailed descriptions of environmental information. 
While captioning model based on our pre-trained audio-text retrieval model shows a significant performance improvement, 
and is able to infer where the sound occurs precisely.
This observation signifies that Auto-ACD showcases an extensive lexicon, enabling the portrayal of a given audio using various sentence structures. On the other hand, it illustrates that models trained on our dataset will deduce the context in which the sound emanates.

\begin{table}[ht]
    \caption{The automatic audio captioning results on Clotho and Auto-ACD test sets. ``S-BERT''  refers to SentenceBERT, ``Env.''  refers to whether the predicted captions contain environmental information.}
    \label{tab:captioning}
    \fontsize{8.5pt}{9.5pt}\selectfont
    
    \setlength{\tabcolsep}{2.5pt} 
    \centering
    \begin{tabular}{ccccccc}
    
     \toprule
    Eval Set & Audio Encoder & Meteor & RougeL & Spider & S-BERT & Env.\\
    \midrule

    \multirow{2}*{Clotho}  &  \multicolumn{1}{c}{CLAP}& 15.5 & 34.9 & 20.6& 46.0 & $\times$  \\
    &  \multicolumn{1}{c}{Ours} & \textbf{16.6} & \textbf{36.2} & \textbf{21.2}  & \textbf{47.4} & $\times$ \\
    \midrule
    
    \multirow{2}*{Auto-ACD}  &  \multicolumn{1}{c}{CLAP}& 9.9 & 23.0 & 19.6 & 8.7 & $\times$ \\
    &  \multicolumn{1}{c}{Ours} & \textbf{21.3} & \textbf{37.9} & \textbf{56.7} & \textbf{10.1} & $\checkmark$\\
    \bottomrule
    \end{tabular}
    
\end{table}

\subsection{Zero-shot Classification}

\subsubsection{Dataset.}
Auto-ACD stands out for integrating its incorporation of environmental information within its text descriptions. Following the training on Auto-ACD, we conduct environmental classification in four distinct scenarios: 
(i) a collection of samples from the AudioSet evaluation set, annotated with child classes of "Acoustic Environment" within the AudioSet ontology, referred to as \textit{AudioSet Env}. To prevent data leakage, here we exclusively utilize the model pre-trained on $\text{Auto-ACD}_\text{VS}$ for this experiment; (ii) the urban acoustic scene dataset~\cite{heittola2020tau}, known as \textit{DCASE 2020 Mobile}, previously utilized in the DCASE 2020 challenge. (iii) the popular urban sound event classification dataset, UrbanSound 8k~\cite{salamon2014dataset}; (iv) the music genre classification dataset, GTZANGenres~\cite{tzanetakis2002musical}.

\subsubsection{Metrics.}
We approach zero-shot classification as an audio-text retrieval experiment, employing a conventional paraphrasing template: "The sound in [environment label] / of [label]." We utilize Recall@1 as the metric for evaluating the environment classification outcomes in this experiment.

\subsubsection{Results.}
The experimental results, as illustrated in Table.~\ref{tab:environment}, highlight the superior environmental recognition capability of ATR pre-trained on Auto-ACD in comparison to CLAP.
Notably, on the AudioSet Env, our model significantly outperforms CLAP, even though we only utilize $\text{Auto-ACD}_\text{VS}$, for pre-training without any data leakage from AudioSet into our training dataset, further serving as a testament to the rich and accurate environmental information in Auto-ACD.  
The results on UrbanSound 8K and GTZANGenres shows that in addition to the audio events, the captions may also include more information, for example, diverse environment descriptions, fine-grained musical genres.

\begin{table}[ht]
    \caption{Zero-Shot Acoustic Environment Classification. ``*" refers to pre-training model on $\text{Auto-ACD}_\text{VS}$. ``US-8K'' refers to UrbanSound 8K.}
    \label{tab:environment}
    \setlength{\tabcolsep}{6pt} 
    \centering
    \begin{tabular}{ccccc}
    \toprule
     Model & AudioSet Env & DCASE  & US-8K & GTZANGenres \\
    \midrule
    CLAP  & 19.5 & 32.2 & 75.0 &  31.5 \\
    Ours & \textbf{39.5}* & \textbf{36.5} & \textbf{76.2} &  \textbf{45.6} \\
    \bottomrule
    \end{tabular}
\end{table}

\section{Conclusion}
In this paper, we present an automatic pipeline for audio caption generation, accompanied by a large-scale and comprehensive audio captioning dataset comprising 1.5M data pairs. Furthermore, we evaluate the performance of various audio-language models on our dataset to authenticate the effectiveness, and provide a manually verified test set along with a benchmark for audio-language tasks. These experimental findings unveil the wealth of information and precise descriptions inherent in our data, facilitating the models to learn more robust audio-language representations.

Owing to the fact that a portion of our dataset originates from VGGSound, procured through an automatic pipeline. The transformation from online videos to precise audio-language pairs has evolved into a thoroughly automated and replicable procedure. Consequently, the acquisition of an expanded corpus of audio-language datasets is now a straightforward endeavour. 
Furthermore, as open-source computer vision models and Large Language Models (LLMs) undergo continuous refinement and advancement, the capacity to extract more precise audio-visual indicators improves, subsequently enhancing the precision of inferences and the quality of paraphrasing the final audio captions. 

\begin{acks}
This work is supported by National Key R\&D Program of China (No.2022ZD0161400).
\end{acks}

\balance
\bibliographystyle{ACM-Reference-Format}
\bibliography{ref}

\clearpage
\section{Dataset Analysis}
In this section, we conduct a more thorough analysis of the proposed dataset, Auto-ACD.
In Section~\ref{subsec:statistics} and Section~\ref{subsec:filtering}, 
we compare Auto-ACD with existing audio-language datasets, 
and discuss the necessity for data filtering.
In Section~\ref{subsec:corpus}, 
we present the distribution of vocabulary. 
In Section~\ref{subsec:Comparison} and Section~\ref{subsec:check}, we compare the captions among Laion-Audio-630K, WavCaps and Auto-ACD, and manually check the quality of a subset of Auto-ACD. In Section~\ref{subsec:visualisation}, we present additional examples from Auto-ACD.

\subsection{Dataset Statistics}
\label{subsec:statistics}
In total, we have collected 1.5 million audio samples, 
each with a duration of 10 seconds, accompanied by one detailed caption. As indicated in Table ~\ref{tab:data_analysis}, in comparison to other datasets, Auto-ACD not only surpasses them significantly in terms of volume, but also contains a longer average sentence length. It stands as the only large-scale dataset that includes environmental information within its descriptions. Laion-Audio-630k may possess a higher vocabulary count, but the majority of its lexicon comprises user-uploaded device information and timestamps, which are irrelevant to the audio content.

\begin{table}[ht]
    \centering
    \caption{Comparation with other audio caption datasets. ``Length'' and ``\# Vocab.'' refer to average length and vocabulary. ``Env.'' and ``Auto.'' refer to environmental information and automatic pipeline, respectively.}
    \setlength{\tabcolsep}{2.25pt} 
    \begin{tabular}{lccccc}
    \toprule
    Dataset  & Quantity & Length & \# Vocab. & Env.  & Auto. \\
    \midrule
    AudioCaps~\cite{kim2019audiocaps}  & 57K  & 8.8 & 5K & $\times$  & $\times$ \\
    Clotho~\cite{drossos2020clotho}    & 30K  & 11.3 & 4K & $\times$  & $\times$ \\
    LAION-Audio-630K~\cite{wu2023large} & 630K & 7.3 & \textbf{311K} & $\times$  & $\checkmark$  \\
    WavCaps~\cite{mei2023wavcaps}      & 400K & 7.8  & 29K & $\times$  & $\checkmark$  \\
    {\bf Auto-ACD (ours)}              & \textbf{1.5M} &  \textbf{18.1} &  22K & \textbf{$\checkmark$} & $\checkmark$ \\
    \bottomrule
    \end{tabular}
    \label{tab:data_analysis}
\end{table}

\subsection{Dataset Filtering}
\label{subsec:filtering}

Our data collection procedure relies on strong audio-visual correspondence. However, many entries within AudioSet contain considerable noise, posing challenges to achieving such coherence, for instance, videos with background music, serene speeches, videos depicting gameplay or software tutorials. Such videos typically only encompass two types of audio events: speech and music. Consequently, the generated captions often contain sparse information and exhibit high error rates. We employ an analysis of audio-visual labels and synchronization model to filter these samples. The specific details of this filtering process are described in Section 3.3 of the main text. 

\begin{figure}[h]
  \centering
  \includegraphics[width=1\linewidth]{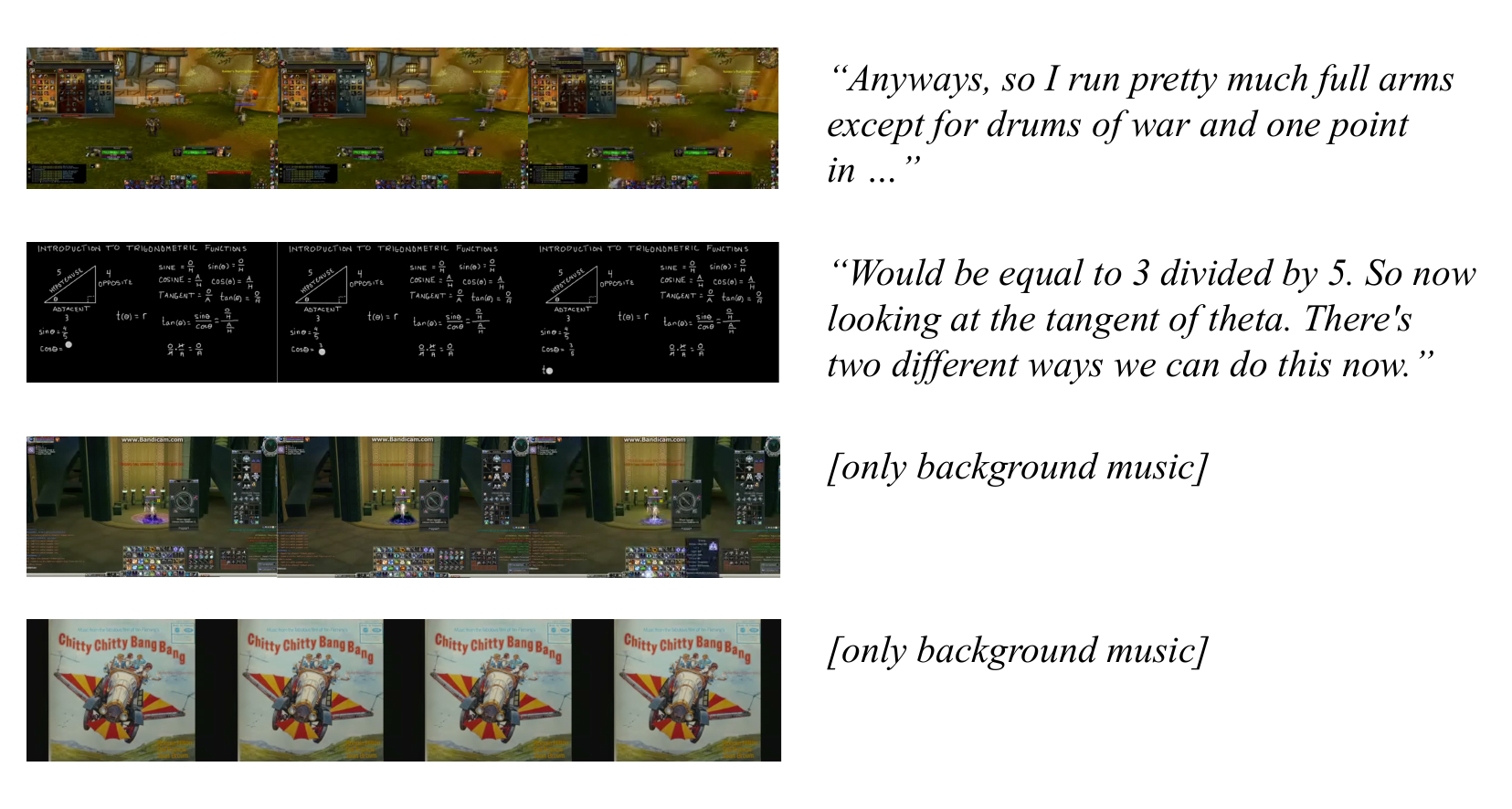}
  \caption{Samples deleted in filter processing. The text on the right side represents transcriptions of speech from the audio in the video, processed using WhisperX.}
  \label{fig:badcase}
\end{figure}

In Figure.~\ref{fig:badcase}, we present some examples of video frame sequences and the outcomes of audio ASR (Automatic Speech Recognition) by WhisperX~\cite{bain2023whisperx} for the excluded data. The majority of discarded entries are due to the audio and video are not unrelated or not synchronized.

\subsection{Dataset Corpus}
\label{subsec:corpus}

\begin{figure}[ht]
  \centering
  \includegraphics[width=1\linewidth]{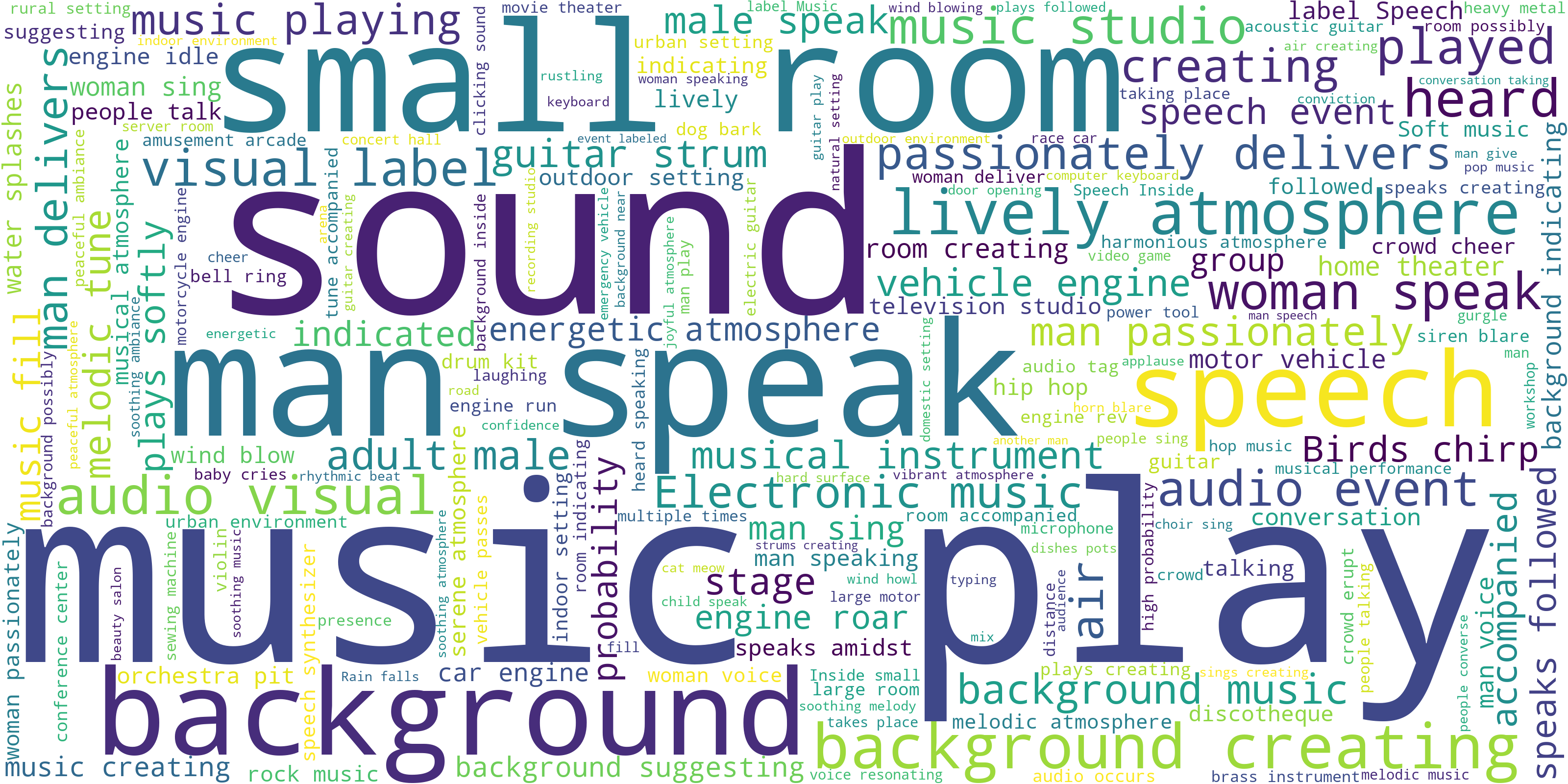}
  \caption{Corpus in Auto-ACD. The higher the frequency of occurrence, the larger the font size of the respective word.}
  \label{fig:wordcloud}
\end{figure}

We visualize the captions for our dataset with word cloud. 
As depicted in Figure.~\ref{fig:wordcloud}, and the common audio tags, \textit{man speak} and \textit{music play} still predominate in frequency within our data. It is noteworthy that terms describing settings, such as \textit{small room} and \textit{music studio}, also emerge with considerable frequency. These are a plethora of audio events, such as \textit{birds chirping}, \textit{engine idling} and \textit{water splashing}, further demonstrating the diverse audio events in Auto-ACD.

\begin{table*}[h]
    \caption{Caption comparison with LAION-Audio-630K and WavCaps, ``LA.'', ``WavC.'' and ``ACD'' refer to LAION-Audio-630K, WavCaps and Auto-ACD, respectively.}
    \label{tab:comparison}
    \setlength{\tabcolsep}{6pt} 
    \centering
    \begin{tabular}{lll}
    \toprule
     No. & Dataset & Generated Caption \\
    \midrule
     \multirow{3}{*}{1.} & LA. & A person is rapping a tree. \\ 
     & WavC. & Music plays with a man rapping. \\
     & ACD. & A woman sings while hip-hop music plays in the background, creating a rapping audio event in a computer room. \\
    \midrule

    \multirow{3}{*}{2.} & LA.& a slushy water lily.\\ 
     & WavC. & Stream noise, crowd and splashing sounds. \\
     & ACD. & A crowd of people yells and cheers as water sloshes in the background at a water park.\\
    \midrule

    \multirow{3}{*}{3.} & LA. & a truck with a siren and a fire engine in an emergency.\\
     & WavC. & A fire engine siren is heard.\\
     & ACD. & An emergency vehicle siren blares loudly as a fire truck rushes through a residential neighbourhood. \\
    \midrule

    \multirow{3}{*}{4.} & LA. & a vehicle with a medium frequency of engine idling. \\ 
     & WavC. & A medium engine sound can be heard.\\
     & ACD. & A medium-sized engine is idling and vibrating, while an adult male speaks in the background near a running vehicle. \\

    \bottomrule
    \end{tabular}
    
\end{table*}

\subsection{Dataset Comparison}
\label{subsec:Comparison}

In Table.~\ref{tab:comparison}, we show example captions from LAION-Audio-630K, WavCaps, and Auto-ACD for the same audio sample. Since the original sounds of the three datasets overlap, we select the descriptions of the same audio in different datasets for comparison.
Specifically, LAION-Audio-630K employs a keyword-to-caption model to transform the tag labels into captions.
WavCaps utilizes ChatGPT to rephrase the tag labels into simple captions. The captions in LAION-Audio-630K and WavCaps tend to be concise, and contain minimal information beyond the audio tags. In particular, the captions of LAION-Audio-630K are short, and may include information deviate from common sense, for instance, ``rapping a tree''.
WavCaps, on the other hand, exhibits a simple sentence structure, such as ``... sound can be heard''.
The captions of these two datasets hardly present more information than audio tags. In contrast, Auto-ACD features longer sentences and provide more comprehensive descriptions of the audio scenes.

\subsection{Dataset Quality Check}
\label{subsec:check}

To evaluate the quality of Auto-ACD, we conduct a manual check on randomly sampled 1000 audio-captions pairs. 
As shown in Table~\ref{manual}, 
(i) we assess the correspondence between our generated captions and original audio; 
(ii) we revise the incorrect words in the final captions and calculate the percentage of modified vocabulary;
(iii) we calculate the ratio of captions that contain inaudible information, for example, colours. 
The results, high correspondence and low erroneous words percentage, indicate that our proposed approach enables high-quality, scalable caption generations, with minimal incorrect information or inaudible information.

\begin{table}[h]
  \caption{Statistics of Manual Check on Auto-ACD.}
  \label{tab:freq}
  \setlength{\tabcolsep}{6.5pt} 
  \begin{tabular}{cccc}
    \toprule
      & Correspondence & Modification & Inaudibility \\
    \cmidrule{2-4} 
    Statistics & 0.924 & 0.053 & 0.042 \\
  \bottomrule
  \label{manual}
\end{tabular}
\end{table}

In addition, we further conduct manual check on each of the steps during caption generation, {\em i.e.}, the various tools used for generating visual clues. We conduct a manual check on randomly sampled 200 audio-captions pairs, to analyse the quality of clues from six different open-source tools and the generated captions. As shown in Table~\ref{tools}, we define a clue that contradicts the audio as incorrect, and we calculate the accuracy of each tool and caption and count the number of correct clues in each sample. 
These tools possess high accuracy, with a high average accuracy at 81.3\% and the highest accuracy at 91.5\%.
we find that 94.0\% of the samples contain at least four correct clues. The fact that 88.0\% of generated captions align with the audio further demonstrates that the LLM is capable of removing incorrect information and producing coherent audio captions.

\subsection{Dataset Visualization}
\label{subsec:visualisation}

As shown in Table.~\ref{tab:visualise}, we show more generated captions for audios from VGGSound and AudioSet. Note that, we present the video sequences to demonstrate how visual information can assist the language description for audio.
It can be observed that, the captions in Auto-ACD not only accurately depict sound events but also infer additional information based on visual priors, that can also be inferred from audios, for example, (i) environmental details, for instance, ``a lively performance arena", ``in a music studio"  and ``a peaceful zen garden", (ii) sound attributes like ``A civil defense siren blares loudly" and ``music plays in the background", (iii) sound variations, for example, ``motorcycle engine revs up and down" and ``a car speeds down a dirt track". 
\begin{table}[h]
  \caption{Accuracy of Manual Check on Open-source Tools. ``Caption.'' refers to AudioCaption model.}
  \label{tab:freq}
  \setlength{\tabcolsep}{3.5pt} 
  \begin{tabular}{ccccccc}
    \toprule
      & BLIP-2 & DINO & CLIP & Place365 & Caption. & PANNs \\
      \cmidrule{2-7}
    Accuracy & 0.915 & 0.755 & 0.805 & 0.725 & 0.770 & 0.905 \\
  \bottomrule
  \label{tools}
\end{tabular}
\end{table}

\begin{table*}[t]
    \caption{Data visualization in Auto-ACD. In each sample, the top line showcases the video frame sequence, the bottom line presents the corresponding audio caption. The sound events in the caption are highlighted in bold text, and environmental information is indicated in \textit{italics} text.}
    \label{tab:visualise}
    \setlength{\tabcolsep}{3pt} 
    \centering
    \begin{tabular}{ll}
    \toprule
     \multirow{2}{*}{No.} & \multirow{2}{*}{Generated Caption} \\
      & \\
    \midrule

    1. & \includegraphics[width=17cm]{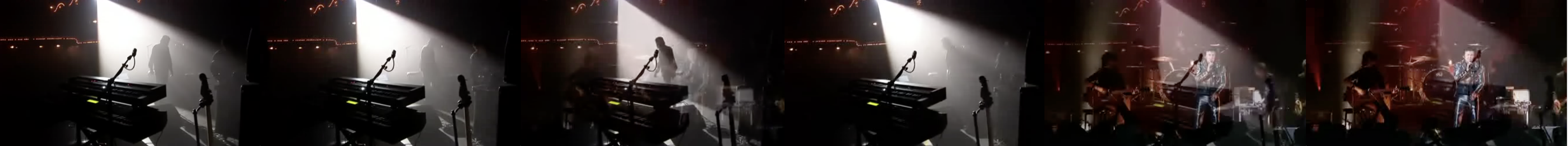}  \\ 
      & \textbf{A man sings} while \textbf{playing the guitar}, accompanied by \textbf{country music} and \textbf{the sound of drums}, \textit{in a music studio}.\\
    \midrule

    2. &  \includegraphics[width=17cm]{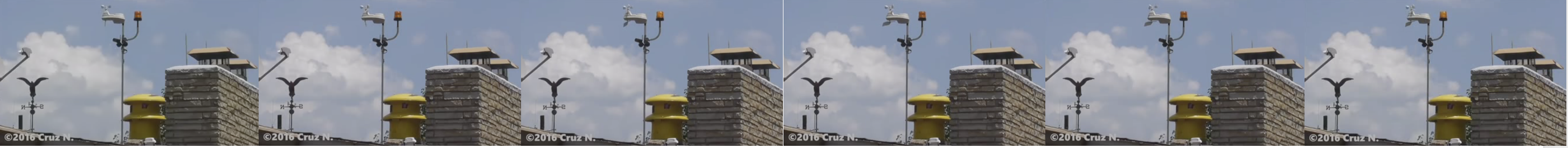}\\ 
      & \textbf{A civil defense siren blares loudly}, indicating an emergency situation, \textit{possibly in a city or urban environment}.\\
    \midrule

    3. & \includegraphics[width=17cm]{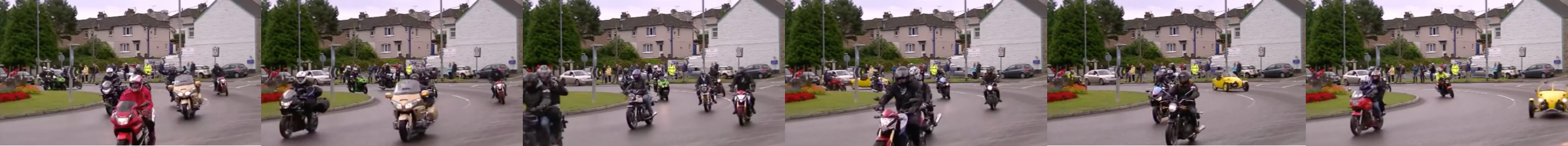} \\ 
      & The \textbf{motorcycle engine revs up and down} while driving through a residential neighborhood, accompanied by \textbf{some speech} and \\
      & \textbf{light engine sounds}.\\
    \midrule

    4. &  \includegraphics[width=17cm]{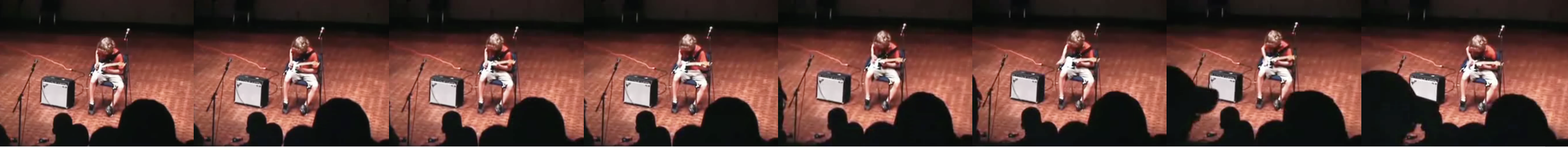}\\ 
      & \textbf{A crowd of people cheer} while \textbf{music plays} in the background, \textit{creating a lively atmosphere in a concert. }\\
    \midrule
    
    5. & \includegraphics[width=17cm]{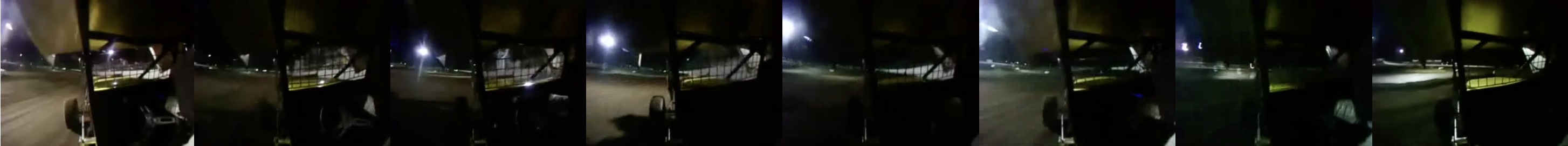} \\ 
      & The sound of a \textbf{loud engine} revving can be heard as \textbf{a car speeds down a dirt track} at night.\\
    \midrule

    6. &  \includegraphics[width=17cm]{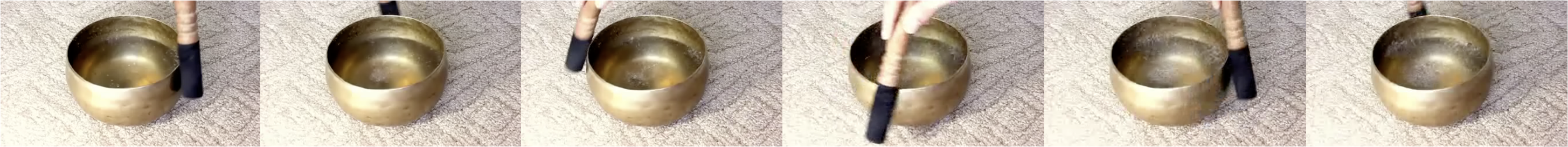}\\ 
      & The sound of \textbf{a singing bowl resonates}, accompanied by \textbf{faint tones of a sine wave and a tuning fork} \textit{in a peaceful zen garden}.\\
    \midrule
       
    7. &  \includegraphics[width=17cm]{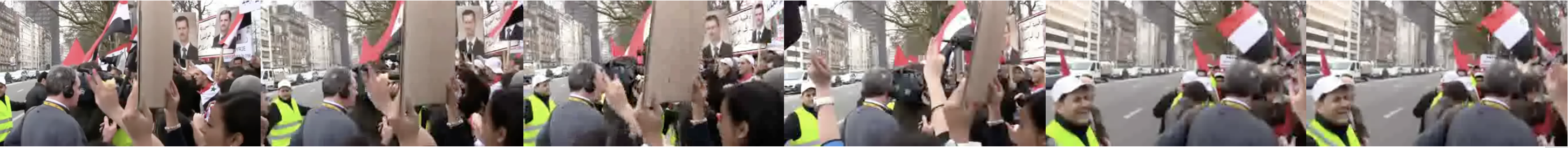}\\ 
      & A group of \textbf{people cheer and sing }while an \textbf{urban battle cry echoes} in the background.\\
    \midrule

    8. &  \includegraphics[width=17cm]{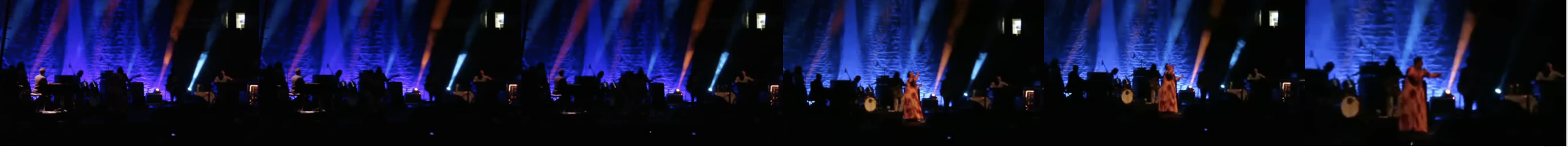}\\ 
      & \textbf{Music plays} as a \textbf{crowd cheers} and \textbf{a band performs} on stage with vibrant lights \textit{in a lively performance arena}.\\

    \bottomrule
    \end{tabular}
    
\end{table*}

\end{document}